\documentclass[prl,twocolumn,showpacs,superscriptaddress,amsmath]{revtex4}
\usepackage{graphicx}
\usepackage{upgreek}
\usepackage{color}

\newcommand{\ket}[1]{\left|{#1}\right>}
\newcommand{\bra}[1]{\left<{#1}\right|}
\newcommand{\tr}[1]{\textnormal{tr}{\left\{#1\right\}}}

\newcommand{\E}{\mathrm{e}}

\begin{document}

\title{Knowledge and ignorance in incomplete quantum state tomography}

\author{Yong Siah Teo}
\affiliation{Centre for Quantum Technologies, National University of Singapore, Singapore 117543, Singapore}
\affiliation{NUS Graduate School for Integrative Sciences and Engineering, Singapore 117597, Singapore}
\author{Huangjun Zhu}%
\affiliation{Centre for Quantum Technologies, National University of Singapore, Singapore 117543, Singapore}
\affiliation{NUS Graduate School for Integrative Sciences and Engineering, Singapore 117597, Singapore}
\author{Berthold-Georg Englert}
\affiliation{Centre for Quantum Technologies, National University of Singapore, Singapore 117543, Singapore}
\affiliation{Department of Physics, National University of Singapore, Singapore 117542, Singapore}
\author{Jaroslav {\v R}eh{\'a}{\v c}ek}
\affiliation{Department of Optics, Palacky University, 17. listopadu 12, 77146 Olomouc, Czech Republic}
\author{Zden{\v e}k Hradil}
\affiliation{Department of Optics, Palacky University, 17. listopadu 12, 77146 Olomouc, Czech Republic}
\pacs{03.65.Ud, 03.65.Wj, 03.67.-a}

\begin{abstract}
Quantum state reconstruction on a finite number of copies of a quantum system with informationally incomplete measurements does, as a rule, not yield a unique result. We derive a reconstruction scheme
where both the likelihood and the von Neumann entropy functionals are maximized
in order to systematically select the most-likely estimator with the largest entropy, that is the least-bias estimator, consistent with a given set of measurement data. This is equivalent to the joint consideration of our partial knowledge and ignorance about the ensemble to reconstruct its identity. An interesting structure of such estimators will also be explored.
\end{abstract}

\date{\today}

\begin{widetext}
\maketitle
\end{widetext}

The acquisition of information about a given ensemble composed of many copies of a quantum system, which is used to identify its quantum state for instance, always depends on the way the measurements are performed. Such a protocol of probing the ensemble is known as quantum tomography. Since our measurement resources are always limited, the information is never complete and the inference of the nature of this ensemble from the measurement data needs to account for the fact that some aspects of the ensemble are fully characterized while others are not.

This is especially the case if the ensemble to be characterized is complex and no feasible set of \emph{informationally complete} measurements is available. In this case, some aspects of the quantum ensemble are not measured and reconstructions of its properties are not unique. For example, the true quantum state of a mode of light made up of an ensemble of photons can be described by a statistical operator $\rho_\text{true}$ residing in an infinite--dimensional Hilbert space, and no matter how ingeniously a measurement scheme is designed to probe this ensemble, an infinite amount of information about it will always remain unknown.

The standard approach to this problem is to apply an \textit{ad hoc} truncation on the Hilbert space and perform the state reconstruction in a particular subspace. This results in a smaller number of unknown parameters that can then be \emph{uniquely} determined by the measurement scheme. Since the truncation is largely based on the experimentalist's intuition about the expected result, that is the true state that describes an ensemble of infinitely many copies of such quantum systems, this cannot be an objective method \cite{rehacek5}. A more objective alternative is to consider the largest possible reconstruction subspace which is compatible with any existing \emph{prior} knowledge about the ensemble. For example, if an experimentalist has prior knowledge of the range of the energy spectrum of a particular quantum ensemble, he should consider the largest possible reconstruction subspace that contains quantum states describing the ensemble in this range of energies. This inevitably introduces more unknown parameters which cannot be unqiuely determined by the measurements and we should select the state estimator in this subspace that is least biased. In this Letter, we show how to carry out this procedure using two celebrated principles --- maximum likelihood (ML)~\cite{fisher,helstrom} and maximum entropy (ME)~\cite{jaynes}. They will be utilized concurrently to yield a unique and objective state reconstruction scheme, the \emph{maximum likelihood--maximum entropy} reconstruction scheme (MLME), a synthesis of knowledge and ignorance.  This scheme will be applicable for any general set of quantum measurements, in particular for those which are non-commuting.

By maximizing the likelihood, information is extracted from the measured data~(\emph{knowledge}) in an optimal way~\cite{helstrom} and an estimator for the true state describing a given ensemble of quantum systems can be reconstructed. However, due to the informational incompleteness of the set of measurements, there exists in general a convex set of estimators which are all maximally likely. We shall select from this set the estimator with the largest entropy, which represents the least-bias guess of the true state consistent with the measurement data. As both the entropy and likelihood functionals are convex, our proposed reconstruction scheme will guarantee a unique solution even for informationally incomplete measurement schemes. This reconstruction procedure minimizes the spurious details coming from the parameters which are not uniquely determined by the measurements.

Given a source that produces identical copies of quantum systems, each in the state
described by the statistical operator $\rho_\text{true}$, one can perform measurements
on $N$ such copies. These measurement outcomes are described by a set of positive operators $\Pi_j$ that compose a POM (probability operator measurement), with $j$ running over all outcomes, and a list of outcome occurrences $n_j$, $\sum_jn_j=N$, or correspondingly the measured frequencies $f_j=\frac{n_j}{N}$, is the raw data obtained from these measurements. Throughout this analysis, we shall assume that the measurements are perfect in the sense of $\sum_j\Pi_j=1$. The next step is to infer the unknown state $\rho_\text{true}$ from the measurement data. For this, we look for estimators --- the ML estimators --- which maximize the \emph{likelihood functional} defined as
\begin{equation}
\mathcal{L}(\rho)=\Bigg(\prod_j p_j^{f_j}\Bigg)^N\,,
\end{equation}
with $p_j=\tr{\rho\Pi_j}$. Using this method, one can always obtain positive estimators suitable for statistical predictions \cite{rehacek1}.

A POM is \emph{informationally complete} if the mapping of statistical operators $\rho$ on the probabilities $p_j$ is injective. This results in a unique operator $\hat{\rho}_\text{ML}\geq 0$ that maximizes $\mathcal{L}(\rho)$, where a hat is used to denote an estimator. We shall focus on the case in which the POM is not informationally complete. This means that there is a convex set of estimators that maximize $\mathcal{L}(\rho)$ for a given set of $f_j$s. The next task is now to choose the one (the MLME estimator) which maximizes the von Neumann entropy \cite{neumann} $S(\rho)=-\tr{\rho\log\rho}$ over the set
of ML estimators. It is well-known that the ME estimator is of the form
\begin{equation}
\hat{\rho}_\text{ME}=\frac{\E^{\sum_j\mu_j\Pi_j}}{\tr{\E^{\sum_j\mu_j\Pi_j}}}\,
\label{mlme_state}
\end{equation}
with real $\mu_js$. Note that our proposed strategy is fundamentally different from the standard ME scheme. The latter involves searching for the ME estimator $\hat\rho_\text{ME}$ which produces the measured frequencies of POM outcomes or moments of a particular observable and maximizing its entropy subjected to these linear constraints \cite{buzek,paris}. Since these constraints can become incompatible with one another due to the presence of statistical noise, this procedure may fail because there simply is no ME estimator.

In our current strategy, instead of taking the data as strict \emph{linear constraints}, information is extracted from the data via the \emph{nonlinear} ML technique and mapped onto the convex subset of quantum states constituting the plateau of the likelihood functional. Thereafter, the entropy is maximized to yield a unique unbiased estimator.
Conceptually, this can be turned into a convex optimization problem of maximizing the objective functional
\begin{equation}
\mathcal{I}(\lambda;\rho)=\lambda S(\rho)+\frac{1}{N}\log\mathcal{L}(\rho)\,,\quad\lambda\geq 0\,,
\label{newinfo}
\end{equation}
which involves a single parameter $\lambda$ as a weight on $S(\rho)$. This $\mathcal{I}(\lambda;\rho)$ is the amalgam of two separate measures of information and so reflects the joint consideration of knowledge and ignorance. In fact, we note that up to an irrelevant additive constant, $\mathcal{I}(\lambda;\rho)$ is a weighted sum of two entropies: $S(\rho)$ which quantifies the \textquotedblleft lack of information\textquotedblright\,and the negative of $S(\{f_j\}|\{p_j\})=\sum_jf_j\log(f_j/p_j)$ (relative entropy) which quantifies the \textquotedblleft gain of information\textquotedblright\,from the measurements. Hence, Eq.~(\ref{newinfo}) is indeed a natural combination of two complementary aspects of information.

Since $\updelta\mathcal{L}(\rho)/\updelta\rho=0$ due to the constraint of maximal likelihood, varying $\mathcal{I}(\lambda;\rho)$ with respect to $\rho$ for \emph{fixed} $\lambda$ gives
\begin{equation}
\frac{\updelta\mathcal{I}(\lambda;\rho)}{\updelta\rho}=\lambda\frac{\updelta S(\rho)}{\updelta\rho}\,.
\end{equation}
In order to maximize $\mathcal{I}(\lambda;\rho)$, we need the derivative to be zero and this is obtained only when $\lambda\rightarrow 0$. In other words, in order to search for the MLME estimator via Eq.~(\ref{newinfo}), it is necessary to take \emph{both} of our knowledge and ignorance of the unknown true state into consideration in such a way that our ignorance takes an infinitesimal weight. Geometrically, the log-likelihood functional is much larger than the entropy functional in this limit and since the log-likelihood functional has a plateau corresponding to the convex set of most-likely estimators, a tiny admixture of the entropy functional introduces a gentle convex hill top \emph{within} the plateau to select the maximum entropy estimator.

By parameterizing a given statistical operator~$\rho$ as $\rho=A^\dagger A/\tr{A^\dagger A}$ to ensure positivity, one can invoke the method of steepest ascent and derive an iterative routine for MLME. Here, we simply mention that one can start from the maximally-mixed state and iterate the equations
\begin{align}
\rho_{k+1}&=\frac{\left(1+\epsilon T_k\right)\rho_k\left(1+\epsilon T_k\right)}{\tr{\left(1+\epsilon T_k\right)\rho_k\left(1+\epsilon T_k\right)}}\,,\label{iter1}\\
T_k&=R_k-1-\lambda\left(\log\rho_k-\tr{\rho_k\log\rho_k}\right)\,
\label{iter2}
\end{align}
with a step size $\epsilon$ and $R_k=\sum_j\left(f_j/p^{(k)}_j\right)\Pi_j$, until the extremal equations~$T_{k'}\rho_{k'}=\rho_{k'}T_{k'}=0$ are satisfied for a particular $k=k'$ with some numerical precision. We denote this operator as the MLME estimator $\hat\rho_\text{MLME}\equiv\rho_{k'}$. If $\lambda=0$, we recover the ML iterative scheme \cite{rehacek1}.

We compare the MLME scheme with the standard ME scheme using the simple example of a trine POM defined by the three outcomes $\Pi_0=(1+\sigma_z)/3$ and $\Pi_\pm=(1\pm\sqrt{3}\sigma_x/2-\sigma_z/2)/3$, where $\sigma_x$ and $\sigma_z$ are standard Pauli operators. A straightforward calculation shows that when $n_0=6$, $n_+=2$ and $n_-=1$ after measuring $N=9$ copies for instance, the standard ME scheme fails as no quantum state has the frequencies $f_0=2/3$, $f_+=2/9$ and $f_-=1/9$ as probabilities. On the other hand, the MLME scheme still gives a positive estimator described by the Bloch vector $(0.194,0,0.981)$ for those frequencies, thus showing its versatility. Only when the frequencies are probabilities giving positive estimators may we use the ME scheme and in this case, the MLME scheme naturally incorporates these constraints.

\begin{figure}[h!]
  \centering
  \includegraphics[width=0.4\textwidth]{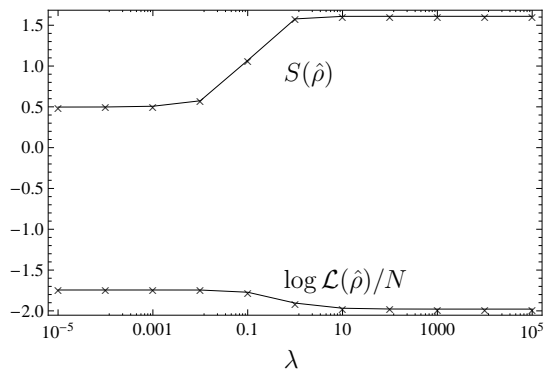}
  \caption{A simulation on quantum tomography on a randomly generated mixed state of light in the five--dimensional Fock space. In this plot, the number of copies of quantum systems measured is fixed at $N=10^4$. A choice of 20 quadrature eigenstates made up of four different $\vartheta$ settings, with five $x$ values corresponding to each setting, which are projected onto this space was used and state estimators are constructed for different values of $\lambda$. As $\lambda$ decreases, both the entropy and likelihood functionals approach their respective optimal values obtained from MLME (i.e. when $\lambda\rightarrow 0$). When $\lambda$ is zero, there is a convex set of estimators giving the optimal likelihood value. For very large $\lambda$ values, the estimators approach the maximally-mixed state and hence $S(\rho)$ approaches the maximal value $\log 5$.}
  \label{fig:behavior}
\end{figure}

To discuss the method of choosing $\lambda$, we shall apply the MLME scheme to homodyne tomography, a technique which is used to reconstruct quantum states of light \cite{homodyne}. This is typically done by measuring a POM which resembles a set of eigenstate projectors $\ket{x_\vartheta}\bra{x_\vartheta}$ of quadrature operators $X\cos\vartheta+P\sin\vartheta$ for various $\vartheta$ values, where $X$ and $P$ are respectively the position and momentum quadrature operators and $x$ and $\vartheta$ are parameters specifying these projectors. It is clear that a finite set of such measurements is never informationally complete in the infinite-dimensional Hilbert space and thus the MLME scheme is necessary to obtain a unique estimator. Fig.~\ref{fig:behavior} shows the dependence of $\log\left(\mathcal{L}(\hat\rho)\right)/N$ and $S(\hat\rho)$ on $\lambda$ such that $\updelta\mathcal{I}(\lambda\rightarrow 0;\hat\rho)=0$. In practice, $\lambda$ can be chosen from a range near zero, within which $\log\left(\mathcal{L}(\hat\rho)\right)/N$ and $S(\hat\rho)$ remain almost constant.

\begin{figure}[h!]
  \centering
  \includegraphics[width=0.4\textwidth]{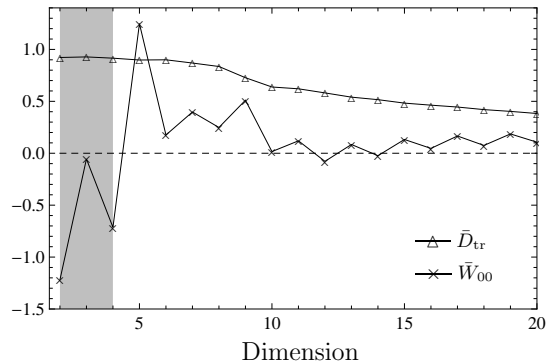}
  \caption{A simulation on quantum tomography on a randomly generated mixed state $\rho_\text{true}$ of light in the 20--dimensional Fock space with a slightly positive $W_{00}=0.141\,$. $\bar D_\text{tr}$ and $\bar W_{00}$ respectively denote the trace-class distance between the reconstructed estimator and the true state and the Wigner function at the phase space origin, both averaged over 50 experiments with $N=10^4$. The same set of 20 quadrature eigenstates as in Fig.~\ref{fig:behavior}, projected onto this space was used and this set of measurements is informationally complete in the \mbox{two--,} three--, and four--dimensional Fock subspaces (shaded region). The values $\bar W_{00}$ and $\bar D_\text{tr}$ were obtained by ML \cite{homodyne} in subspaces of dimensions two to four, and by the MLME scheme in dimensions greater than four. The plot shows a strong dependence of $\bar W_{00}$ and $\bar D_\text{tr}$ on the subspace dimension. In this case, it is obvious that the negativity of $\bar W_{00}$ inferred by a reconstruction in a subspace too small is just an artifact of the truncations. Also, $\bar D_\text{tr}$ decreases as the reconstruction subspace increases in dimension. This demonstrates the advantages of the MLME scheme over the ML method.}
  \label{fig:negativity}
\end{figure}

Homodyne tomography is commonly used not only in quantum tomography on the true state, but also in quantum diagnostics where a given true state is to be classified as being classical/non-classical or separable/entangled. A typical quantity which is often investigated as an indication of whether an unknown true state is non-classical is the value of the Wigner functional at the phase space origin evaluated with a reconstructed estimator $\hat\rho$ for the unknown true state. This is defined as $W_{00}=2\tr{\hat\rho \mathcal{P}}$, with $\mathcal{P}=\int\mathrm{d}x\,\ket{x_\vartheta}\bra{-x_\vartheta}$ for any $\vartheta$. In particular, a negative value for $W_{00}$ implies that $\hat\rho$ is a non-classical state. To obtain an estimator $\hat\rho$, one would need to choose a subspace from the infinite-dimensional Hilbert space in which the reconstruction procedure is tractable. This means that the value of $W_{00}$ will depend on this truncation, which in turn relies on the prior knowledge one has about the true state. Using our scheme, we perform a simulation, shown in Fig.~\ref{fig:negativity}, to illustrate this dependence.

If the true state lies outside the subspace of interest, then the estimated value of $W_{00}$ can drastically deviate from the true value. It is clear that a truncation of the Hilbert space into a smaller reconstruction subspace can lead to diagnostics which are highly incompatible with the true result. So, if one is interested in performing an objective quantum tomography experiment on a given ensemble of quantum systems with some prior knowledge regarding its true state, an option would be to reconstruct the MLME estimator in the largest possible subspace based on this prior knowledge. By enlarging the reconstruction subspace, many more admissible states are taken into consideration and more reliable state estimations and quantum diagnostics can thus be performed. We now have an operational reconstruction scheme that combines our knowledge and ignorance about the unknown true state to give us a unique state estimator in an objective way.


There exists an interesting structure in these MLME estimators and to explore it, one needs some knowledge on the structure of the POM used and its influence on the $D$--dimensional Hilbert space. Suppose the set of $K$ POM elements $\Pi_j$ are informationally incomplete. A consequence of this is that the number of linearly independent $\Pi_j$s is less than $D^2$. To determine their linear independence, we can look for the eigenvalues of the $K\times K$ Gram matrix $M$ with the matrix elements defined as
\begin{equation}
M_{jk}=\tr{\Pi_j\Pi_k}\,.
\end{equation}
The number of positive eigenvalues $n_{>0}\leq D^2$ of $M$ determines the number of linearly independent measurement outcomes. The largest value of $n_{>0}$ is $D^2$ since this is the maximum number of linearly independent operators spanning the space of Hermitian operators as a basis. Hence a set of informationally incomplete $\Pi_j$s acting on the $D$--dimensional Hilbert space is such that $n_{>0}< D^2$.

Any $D$--dimensional positive operator can be represented by a set of $D^2$ Hermitian basis operators $\Gamma_j=\Gamma^\dagger_j$ satisfying the trace-orthonormality condition $\tr{\Gamma_j\Gamma_k}=\delta_{jk}$. For dimension two, an example of such a basis is the the familiar set of four operators $1/\sqrt{2}$,  $\sigma_x/\sqrt{2}$, $\sigma_y/\sqrt{2}$ and $\sigma_z/\sqrt{2}$. Once the number of independent measurement outcomes $n_{>0}$ is known, one can construct a set $\{\Gamma_j\}_{j=1}^{n_{>0}}$ of $n_{>0}$ trace-orthonormal Hermitian basis operators directly from the $K$ POM elements. In other words, each of the $K$ POM elements can be expressed as a linear combination of the $n_{>0}$ basis operators
\begin{equation}
\Pi_j=\sum^{n_{>0}}_{k=1}a_{jk}\Gamma_k\,,
\label{pom_recon}
\end{equation}
where all coefficients $a_{jk}$ are real. This implies that the $n_{>0}$--dimensional subspace is spanned by the basis operators that \emph{uniquely} specify the POM outcomes. We will coin this subspace the \emph{measurement subspace}. The rest of the $D^2-n_{>0}$ Hermitian basis operators, which are trace-orthonormal to the previous set and span the subspace, that is complement to the measurement subspace can also be constructed.

Suppose a state estimator $\hat{\rho}_\text{ML}$ is generated using the ML procedure on a set of measurement data obtained from the POM outcomes $\Pi_j$. We can represent this estimator by a set of Hermitian trace-orthonormal basis operators inasmuch as
\begin{equation}
\hat{\rho}_\text{ML}=\underbrace{\sum^{n_{>0}}_{k=1}c^\text{ML}_k\Gamma_k}_{\equiv\tilde\rho_\text{ML}}+\underbrace{\sum^{D^2}_{k=n_{>0}+1}c^\text{ME}_k\Gamma_k}_{\equiv\tilde\rho_\text{ME}}\,.
\label{rho_decomp}
\end{equation}
The part $\tilde\rho_\text{ML}$ resides in the measurement subspace, which is spanned by the measurement outcomes $\Pi_j$ giving the measurement data, and is uniquely fixed for all ML estimators by the ML procedure for the same set of measurement data. The part $\tilde{\rho}_{\mathrm{ME}}$ resides in the complementary subspace, which is orthogonal to the measurement subspace, and thus does not contribute to the $p_j$s. In other words, $\tr{\tilde\rho_\text{ME}\Pi_j}=0$ and this can imply the existence of a family of $\tilde\rho_\text{ME}$s that gives the same set of ML probabilities as long as the $\hat{\rho}_\text{ML}$s are positive.

Therefore, the MLME scheme can be understood as an optimization over the complementary subspace to maximize $S(\rho)$ under the constraint $\hat{\rho}_\text{MLME}\geq 0$. However, one notes that only certain sets of $c^\text{ME}_j$s are allowed during the optimization due to this positivity constraint. This is especially important when $\hat{\rho}_\text{MLME}$ is rank deficient and lies on the boundary of the state space. Geometrically, the plateau of most-likely states is generally a much smaller subspace contained in the complementary subspace. In some cases, this plateau contains a \emph{single} ML estimator due to the positivity constraint even when the measurements are informationally incomplete. In general, the boundary of the plateau is complicated and deserves further study.

In summary, we have developed a state reconstruction scheme which is applicable to any set of measurements, particularly those which are informationally incomplete, as in homodyne tomography for instance. We emphasize that in order to carry out least-bias state reconstructions and quantum diagnostics on an ensemble of quantum systems, a good way is to do this over a large subspace of states compatible with some prior knowledge on the ensemble in order to avoid inaccurate results which can have detrimental effects on statistical predictions. We then explored the geometrical structure of the state estimators obtained from this scheme and gave an alternative understanding to the state reconstruction procedure.

This work is supported by NUS Graduate School (NGS) and the Centre for Quantum Technologies which is a Research Centre of Excellence funded by Ministry of Education and National Research Foundation of Singapore, as well as the Czech Ministry of Education,
Project MSM6198959213, and the Czech Ministry of Industry and Trade, Project FR-TI1/364.


\begin{thebibliography}{9}

\bibitem{rehacek5}
J. {\v R}eh{\'a}{\v c}ek, D. Mogilevtsev, and Z. Hradil, New J. Phys. \textbf{10}, 043022 (2008).

\bibitem{fisher} R. A. Fisher, Phil. Trans. R. Soc. London \textbf{A 222}, 309 (1922).

\bibitem{helstrom}
C. W. Helstr{\o}m, {\it Quantum Detection and Estimation
Theory}, Academic Press, New York (1976).

\bibitem{jaynes}
E. T. Jaynes, Phys. Rev. \textbf{106}, 620 (1957), Phys. Rev. \textbf{108}, 171 (1957).

\bibitem{rehacek1}
J. {\v R}eh{\'a}{\v c}ek, Z. Hradil, E. Knill, and A. I. Lvovsky, Phys. Rev. A \textbf{75}, 042108 (2007).

\bibitem{neumann} J. von Neumann, \emph{Mathematical Foundations of Quantum Mechanics} (Princeton University Press, 1955).

\bibitem{buzek} V. Bu{\v z}ek, G. Adam, and G. Drobn{\'y}, Ann. Phys. (N.Y.) \textbf{245}, 37 (1996).

\bibitem{paris}
A. R. Rossi, and M.G.A. Paris, Eur. Phys. J. D \textbf{32}, 223 (2005).

\bibitem{homodyne} D. T. Smithey, M. Beck, M. G. Raymer, and A. Faridani, Phys.
Rev. Lett. \textbf{70}, 1244 (1993); A.~Ourjoumtsev, R. Tualle-Brouri, P.~Grangier, Phys.
Rev. Lett. \textbf{96}, 213601 (2006); J. S. Neergaard-Nielsen, B. M. Nielsen, C. Hettich,
K. M{\o}lmer, E. S. Polzik, Phys. Rev. Lett. \textbf{97}, 083604 (2006).

\end{thebibliography}
\end{document}